\documentstyle[12pt]{article}
\def\x{{\mbox{\boldmath$x$}}}
\def\u{{\mbox{\boldmath$u$}}}
\def\uu{{\mbox{\boldmath$u$}}}
\def\r{{\mbox{\boldmath$r$}}}

\def\u1rms{u_{1,rms}}
\def\eps{\epsilon}
\def\ceps{c_\epsilon}
\def\cepsi{c_{\epsilon  ,\infty}}
\def\begineq{\begin{equation}}
\def\endeq{\end{equation}}

\begin{document}
\title{Crossover from high to low Reynolds number turbulence}
\author{Detlef Lohse}
\maketitle

\centerline{The James Franck Institute, The University of Chicago,}
\centerline{ 5640 South Ellis Avenue, Chicago, IL 60637, USA}

\bigskip

\date{}
\maketitle
\bigskip
\bigskip
\bigskip

The Taylor-Reynolds and Reynolds number ($Re_\lambda$ and $Re$)
dependence of the dimensionless energy dissipation rate
$\ceps =\eps L / \u1rms^3$ is derived for statistically stationary
isotropic turbulence, employing the results of a variable range
mean field theory. Here, $\eps$ is the energy dissipation rate,
$L$ is the (fixed) outer length scale, and $\u1rms$ a rms velocity
component. Our fit-parameter free results for $\ceps (Re_\lambda)$
and also for $Re_\lambda (Re)$ are in good agreement with
experimental data. Using the $Re$-dependence of $\ceps$
we account for the time dependence of the mean vorticity
$\omega (t)$ for decaying isotropic turbulence, which
was recently experimentally examined [M.\ Smith, R.\ J.\
Donelly, N.\ Goldenfeld,
and W.\ F.\ Vinen,
Phys.\ Rev.\ Lett.\ 71, 2583 (1993)].
The lifetime of decaying turbulence, depending on the
initial $Re_{\lambda ,0}$, is predicted and found to saturate
at $0.18 L^2/\nu \propto Re_{\lambda ,0}^2$ ($\nu$ is the viscosity)
for large $Re_{\lambda ,0}$.
\vspace{0.5cm}\noindent


\newpage
Dimensional analysis has proven to be a powerful tool in turbulence
research, giving a number of
key features of turbulent spectra \cite{kol41,ll87}.
The main idea is to connect small scale quantities such as the energy
dissipation rate $\eps$ with the
large scale quantities such as the outer length
scale $L$ and a rms velocity component, $\u1rms$. More precisely,
following Richardson's cascade picture of turbulence \cite{ric26,ll87},
it is argued that for fully developed turbulence
\begineq
\eps =\ceps \u1rms^3 /L
\label{eq1}
\endeq
holds, where $\ceps$ is a dimensionless constant in the range of 1.

However, it is not clear, for what Reynolds number $Re$
the large $Re$ limit (\ref{eq1}) is reached.
For small $Re$ the dimensionless number $\ceps$ clearly
depends on $Re$. E.g., for plane
Couette flow with shear $2\u1rms/L$ in one
direction we have $\ceps (Re) = 4/Re$. Here and henceforth,
following Sreenivasan \cite{sre84}, we defined the Reynolds number
$Re$ as
\begineq
Re=L\u1rms /\nu,
\label{eq2}
\endeq
where $\nu$ is the viscosity. Even for large $Re$ it is not clear whether
the $Re$-dependence of $\ceps$ vanishes. Recent experiments on the Taylor
Couette flow, which is {\it bounded}, seem to show that a logarithmic
$Re$-dependence persists up to very high $Re$ \cite{lat92}.
(Here, $Re$ is defined with the velocity of the outer
rotating cylinder.) It can be
accounted for by employing Prandtl's boundary layer theory
\cite{ll87,pra25,doe94},
yet mathematically only $\ceps < const$ can be proven
up to now \cite{doe92,doe94}.
For {\it unbounded} flow experiments favor $\ceps = const$ for high $Re$,
see Sreenivasan's collection of data for grid turbulence \cite{sre84}.

Another way of expressing the physical contents of (\ref{eq1}) is to
give the $Re$-dependence of the Taylor-Reynolds number $Re_\lambda=
\lambda \u1rms /\nu$, where $\lambda = \u1rms / (\partial_1u_1)_{rms}$
is the Taylor microscale. With $\eps = 15 \nu (\partial_1u_1)^2_{rms}$,
valid for isotropic turbulence \cite{ll87}, and eqs.\ (\ref{eq1}),
(\ref{eq2}) we get
\begineq
Re_\lambda = \sqrt{15 Re/ \ceps (Re)}.
\label{eq3}
\endeq
If we have $\ceps = const$ for large $Re$, we thus also have
$Re_\lambda \propto \sqrt{Re}$.

In this Letter we will first derive  {\it explicit}
expressions
for $\ceps (Re)$ and $\ceps (Re_\lambda )$ for unbounded flow.
They only depend on the Kolmogorov constant $b$, which is
{\it known} from many experiments to be $b=8.4$ \cite{my75}.
We do not introduce any free parameter.
This is possible as we employ the results from the variable range mean
field theory \cite{eff87}, which embodies the Navier-Stokes
{\it dynamics}.
We thus offer a way to
go far beyond dimensional analysis. In the second part of the
paper we apply our results to decaying turbulence, which has very recently
been experimentally examined and analyzed by dimensional analysis
by Smith, Donelly, Goldenfeld, and Vinen
\cite{smi93a}. Both our results for $\ceps (Re_\lambda )$
and for the time dependence of the mean vorticity $\omega (t)$ in decaying
turbulence are in good agreement with experimental data.

We start from the final result of Effinger and Grossmann's
variable range mean field theory \cite{eff87}. For homogeneous,
isotropic turbulence a differential equation
for the velocity structure function
$D(r) = \langle (\uu(\x + \r ) - \uu( \x ))^2
\rangle$ is derived \cite{eff87}, namely
\begineq
\eps = {3\over 2}\left( \nu + {D^2(r)\over b^3 \eps}\right)
{1\over r}{d\over dr} D(r).
\label{eq4}
\endeq
For small $r$ the solution
is $D(r)= \eps r^2 / (3\nu )$, whereas for large $r$
we get the well known Kolmogorov result $D(r) = b (\eps r)^{2/3}$. The
Kolmogorov constant $b$ can be calculated within the approach of
ref.\ \cite{eff87}. But rather than taking the mean field result
$b=6.3$ \cite{eff87}, we
will use eq.\ (\ref{eq4}) with the experimental value
$b=8.4$ \cite{my75}, as this value gives better agreement
with the measured structure function $D(r)$, see Fig.\ 4 of ref.\
\cite{eff87}.
Eq.\ (\ref{eq4}) is an energy balance equation.
The first term in the brackets corresponds to viscous dissipation,
the second one can be interpreted as eddy viscosity.

Integrating the differential equation (\ref{eq4})
from 0 to the outer length scale $L$ we get
\begineq
D^3(L) + 3b^3\eps\nu D(L) -b^3 \eps^2 L^2 =0.
\label{eq5}
\endeq
Here we have assumed that (\ref{eq4}) holds for all $r$ up to $r=L$,
which is definitely not the case for bounded flow and even for
unbounded flow slight corrections might be necessary, see below.
Anyhow, if $L$ is large enough and the flow is isotropic,
it is $D(L) = 2 \langle {\uu}^2 \rangle = 6\u1rms^2$. With eqs.\
(\ref{eq1}) and (\ref{eq2})
the quadratic (in $\eps$) equation (\ref{eq5}) can be written in
dimensionless form as
\begineq
\ceps^2 - {18\over Re} \ceps -\left( {6\over b}\right)^3=0,
\label{eq6}
\endeq
which is easily solved to give
\begineq
\ceps (Re)=\cepsi \left\{ \left({3b^3\over 8}\right)^{1/2} {1\over Re}
+ \sqrt{1+ {3b^3\over 8} {1\over Re^2}} \right\}.
\label{eq7}
\endeq
The dependence of $\ceps$ on the {\it Taylor-}Reynolds number $Re_\lambda$
can be obtained from eqs.\ (\ref{eq6}) and (\ref{eq3}),
\begineq
\ceps (Re_\lambda) = \cepsi \sqrt{1+ {5\over 4} {b^3\over Re^2_\lambda}}.
\label{eq8}
\endeq
In both  formulae we have introduced $\cepsi = (6/b)^{3/2} = 0.60$
only for {\it convenience}; the $Re$- and $Re_\lambda$- dependences
are purely determined by the Kolmogorov constant $b$.

For large $Re$ or $Re_\lambda$, the function
$\ceps (Re_\lambda)$ indeed becomes {\it constant},
$\ceps (Re_\lambda ) = \cepsi = (6/b)^{3/2}$, as Sreenivasan finds for
grid turbulence with biplane square mesh grids \cite{sre84}. The
experimental value $\cepsi \approx  1.0$ is slightly larger
than our result $\cepsi = 0.60$. The reason
for the slight discrepancy is likely
due to non universal boundary effects. We
have assumed (\ref{eq4}) to hold for all $r$ up to $r=L$, thus
$D(L) = 6 \u1rms^2 = b(\eps L)^{2/3}$, which already results in
$\cepsi = (6/b)^{3/2} = 0.60$. Yet for $r$ around $L$, $D(r)$ will
not scale as $D(r) = b(\eps r )^{2/3}$ and we have
$D(L) < b (\eps L)^{2/3}$.
This non universal boundary effect might be treated by introducing
an effective, geometry depending $b^{(eff)} < b$
instead of $b$ in eq.\ (\ref{eq5}),
defined by $D(L)=b^{(eff)}(\eps L)^{2/3}$. (Note that an introduction
of $b^{(eff)}$ already in (\ref{eq4}) is inappropriate, as boundary
effects should not be seen in $D(r)$
for small $r$.) To get the experimental
$\cepsi = 1.0$, one should have $b^{(eff)}= 6/\cepsi^{2/3} =6< b=8.4$.

The quantity $\ceps (Re_\lambda)/\cepsi $, eq.\ (\ref{eq8}), is
plotted in Fig.\ 1, together with Sreenivasan's experimental data
for grid turbulence.
For $Re_\lambda\approx 50$ the function $\ceps (Re_\lambda )$
saturates at $\cepsi$, in good agreement with the data.

For really small $Re$ eq.\ (\ref{eq7}) can -- strictly speaking --
no longer be applied, as laminar flow is never isotropic, whereas
eq.\ (\ref{eq4}) only holds for turbulent,
isotropic flow.
Note that Fig.\ 1, starting with $Re_\lambda =5$, does not
include the laminar case (as seen from the inset), since in
laminar flow $Re_\lambda$ looses its meaning.
If we perform
the small $Re$ limit
$Re \ll \sqrt{3b^3/8} = 14.9$, $Re_\lambda \ll \sqrt{5b^3/4}=27.2$,
nevertheless, we can get $\ceps (Re) = 18/Re$,
{\it independent} of $b$, as expected, since $b$ characterizes
the highly turbulent state. The $\propto Re^{-1}$-dependence is
correct. The prefactor 18 is -- again as expected -- too large,
if compared to the highly anisotropic laminar
Couette flow with shear in only {\it one} direction, see above. In
more isotropic laminar flow, e.g., in flow
with shear in {\it three}
directions, the  agreement for small $Re$ will be better.

Equations (\ref{eq3}) and (\ref{eq7}) give the function $Re_\lambda(Re)$,
see inset of Fig.\ 1, which strongly resembles latest experimental
measurements for grid turbulence by Castaing and Gagne
\cite{cas93b,cas93p}.
Also Grossmann and Lohse obtain a very similar curve $Re_\lambda(Re)$
from a reduced wave vector set approximation of the Navier-Stokes
equations \cite{gnlo94b}. Here, for large
$Re$ we have $Re_\lambda = \sqrt{15(b/6)^{3/2}} \sqrt{Re} \approx
5 \sqrt{Re}$. For small $Re$ it is $Re_\lambda = \sqrt{5/6} Re$.
Extrapolating these two limiting cases, the crossover between
them takes place at $Re_{CO} = 18 (b/6)^{3/2} = 29.8$, corresponding
(via eq.\ (\ref{eq3})) to $Re_{\lambda, CO} = 21.6$, which seems
very realistic to us, cf.\ also \cite{cas93p}.

Up to now we applied our theory to statistically {\it stationary}
turbulence. But it also offers an opportunity to analyze
{\it decaying} turbulence.
Most experiments on decaying turbulence have been performed
in wind tunnels up to now, where the distance from the grid
gives the decay time $t$, if the mean velocity is known
\cite{bat53}. In this kind of experiment the outer length
scale $L$ grows with time $t$,
as the wake behind the grid becomes wider with increasing distance.
Yet in a very recent new type of
experiment performed by Smith, Donelly, Goldenfeld, and Donelly
\cite{smi93a}, $L$ can be kept fixed. In that experiment
a towed grid generates homogeneous turbulence in a channel filled
with helium II. The decay of the mean vorticity $\omega (t)$ is
measured by second sound attenuation \cite{smi93a,don91}.
As in \cite{smi93a} we assume that Navier-Stokes dynamics
can be applied to this fluid, see the discussion in \cite{smi93a}.

In the theoretical analysis of this experiment, for very high
$Re_0 = Re(t=0)$,
$Re_{\lambda ,0} = Re_{\lambda}(t=0)$,
the quantity $\ceps$ can be considered constant.
But the smaller $Re(t)$ or
$Re_\lambda (t)$ become with increasing time $t$,
the more important are the corrections seen in eqs.\
(\ref{eq7}),
(\ref{eq8}).
The total energy (per volume)
of the flow is $E= {\uu}^2_{rms}/2 = 3 \u1rms^2/2$. The
decay of the fully developed turbulence is governed by the differential
equation
\begineq
\eps = \ceps {\u1rms^3\over L} = \ceps (E(t)) \left( {2\over 3}\right)^{3/2}
{E^{3/2}(t) \over L} = - \dot E(t).
\label{eq9}
\endeq
The outer length scale $L$ is fixed, as in the experiment
we are referring to
\cite{smi93a}. We change variables to $Re(t) = \sqrt{2/3} L \sqrt{E(t)}
/\nu$ and obtain
\begineq
\dot Re = -{1\over 3} {\nu \over L^2} \ceps (Re) Re^2.
\label{eq10}
\endeq
Integrating eq.\ (\ref{eq10}) with the initial condition
$Re(t=0) = Re_0$ gives the time dependence $Re(t)$ of the
Reynolds number,
\begineq
{t\over \tau} = {3 \over \cepsi} \int_{Re_0}^{Re(t)}
{ -dx\over \gamma x + x \sqrt{x^2 + \gamma^2}}.
\label{eq11}
\endeq
Here, for simplicity we have introduced the viscous time scale
$\tau = L^2/\nu$ and the constant  $\gamma = \sqrt{3b^3/8}=
9/\cepsi = 14.9$.
The integral can be calculated analytically. We define the
indefinite integral as
\begineq
F(Re) = {1\over 2Re^2} \left\{ -\gamma + \sqrt{\gamma^2 + Re^2} \right\}
+ {1\over 2\gamma} \log
\left\{ {\gamma + \sqrt{\gamma^2 + Re^2} \over Re} \right\}.
\label{eq12}
\endeq
Thus the time dependence of $Re(t)$ is given by the inverse function
of
\begineq
t(Re)/ \tau = 3  (F(Re) -F(Re_0))/\cepsi.
\label{eq13}
\endeq
Note again that there is no free parameter in eq.\ (\ref{eq13}), all
quantities on the rhs can be expressed
in terms of the Kolmogorov constant $b$.

Imagine now the limiting case of large $Re_0$ and also
large time, but
$Re(t) < Re_0$ still large, i.e., $t$ not too large.
For large $Re$ both terms in (\ref{eq12}) contribute
$(2Re)^{-1}[1+O(\gamma /Re)]$.
Thus $F(Re) = 1/ Re$ and from (\ref{eq13}) we explicitly get
\begineq
Re (t)= \left( {1\over Re_0}
+ {\cepsi \over 3} {t\over \tau} \right)^{-1}
\approx {3\over \cepsi} \left({t\over \tau}\right)^{-1}.
\label{eq14}
\endeq
In Fig.\ 2 we plotted $Re_\lambda (t)$,
calculated from eqs.\
(\ref{eq13}) and (\ref{eq3}), for several $Re_0$. The scaling law
$Re_\lambda (t) \approx 3\sqrt{5}
\left( t / \tau \right)^{-1} / \cepsi$
(corresponding to (\ref{eq14})) only starts
to be observable for $Re_0 \approx 10^3$, i.e., $Re_{\lambda,0}
\approx 156$.

In the final period of decay, i.e., for very large $t$ (large enough
so that $Re(t) \ll \gamma$), we get $Re(t)=2\gamma \sqrt{e}
\exp{(-6t/\tau)}$ and $E(t)=(9/4)eb^3\nu^2 L^{-2} \exp{(-12t/\tau)}$.
An exponential decay for very large $t$
also holds for decaying turbulence with growing
outer length scale $L(t)$ \cite{bat53}.

To compare our results with the helium II experiment \cite{smi93a},
we have to calculate the mean vorticity $\omega (t)$.
Vorticity always causes strain in the flow. It can be shown \cite{ll87} that
$\nu \omega^2 = \eps$.
Thus
\begineq
\nu \omega^2 = \eps = - \dot E = - 3\nu^2 Re \dot Re /L^2.
\label{eq15}
\endeq
With eq.\ (\ref{eq10}) we get
\begineq
\tau\omega (t) = Re \sqrt{Re \ceps (Re)} = \sqrt{\cepsi} Re(t)
\sqrt{\gamma + \sqrt{\gamma^2 + Re^2(t)}},
\label{eq16}
\endeq
where the universal law on the rhs again only depends on the Kolmogorov
constant $b$ and on the time $t$.

Next, we estimate the lifetime $t_l$ of the decaying turbulence.
For this purpose  we
calculate, how the Kolmogorov length \cite{ll87}
\begineq
\eta (t) = (\nu^3/\eps (t))^{1/4}=
\sqrt{\nu/\omega (t)},
\label{eq17}
\endeq
depends on time $t$.
Of course, $\eta (t)$ will increase with time, as the turbulence becomes
weaker and weaker. The behavior
can be obtained from eqs.\
(\ref{eq16}) and (\ref{eq13}) for any $Re_0$. If $Re_0$ is large enough,
scaling $\eta (t)/L \propto (t/\tau)^{3/4}$ can develop.

How to define the lifetime $t_l$ of the turbulence?
As the crossover
in the structure function
$D(r)$ between the viscous subrange and the inertial
subrange happens at $r\approx 10\eta$ \cite{eff87}, we define
the lifetime
$t_l$ by the condition $10\eta (t_l) = L$. A Reynolds number $Re_l$
is associated with this time $t_l$ via the eqs.\
(\ref{eq16}) and
(\ref{eq17}).
We calculate $Re_l = 20.3$, $Re_{\lambda ,l} = 16.0$, which is,
as it should be, near to the viscous-turbulent
crossover in the curve
$Re_\lambda (Re)$, which occurs at $Re_{CO}=29.8$,
$Re_{\lambda , CO}=21.6$,
see inset of Fig.\ 1.
With this definition we obtain the lifetime $t_l$ of the
decaying turbulence for any given $Re_0$ (or, via (\ref{eq3}),
$Re_{\lambda ,0}$) as
\begineq
t_l(Re_0)/ \tau = 3  (F(Re_l) -F(Re_0))/\cepsi.
\label{eq18}
\endeq
We plotted $t_l (Re_{\lambda ,0})$ in the
inset of Fig.\ 2. For small $Re_{\lambda , 0}$
the lifetime
$t_l (Re_{\lambda ,0})$ grows logarithmically with $Re_{\lambda ,0}$.
For very large $Re_{\lambda ,0}$
it saturates at $t_l(Re_{\lambda ,0}) = 3\tau F(Re_l)/\cepsi
=0.18 \tau $, i.e., it becomes independent on $Re_0$, if measured in
time units  of $\tau = L^2 /\nu$. The lifetime, if measured in seconds,
of course increases $\propto Re_0 \propto Re_{\lambda , 0}^2$ in the
limiting case.

Finally we compare with the  data of
the helium II experiment \cite{smi93a}.
First we have to
embody the boundary effects, as they should be larger in the helium II
experiment than in grid turbulence,
because the turbulence decays in a tube.
Indeed, in ref.\ \cite{smi93a} $\cepsi^{(eff)}
= 36.4$ is given (using our
definition of $\ceps$, eq.\ (\ref{eq1})), corresponding to
$b^{(eff)} = 6/ (\cepsi^{(eff)})^{2/3} = 0.55$.

Using  this $b^{(eff)}$ instead of $b$ (or $\gamma^{(eff)}$,
$\cepsi^{(eff)}$, respectively)
in eqs.\ (\ref{eq12})
and (\ref{eq16}),
we  plotted $\omega(t)$ in Fig.\ 3, together with Smith et al.'s
experimental
data \cite{smi93a}.
The two curves show the same features.
For small $t$ there is no power law.
For medium $t$ the theory gives
$\tau\omega(t) \approx 3^{3/2} (t/\tau )^{-3/2}/\cepsi^{(eff)}$. The power
law exponent $-3/2$, which is clearly seen in the experimental data,
has already been derived by dimensional analysis
\cite{smi93a}. But in theory
the $-3/2$-power law for $\omega (t)$ extends much further
than observable in experiment, because there experimental noise
hinders observation already for $t>10s$.
For further comparison a reduction of the experimental noise is
essential. Experiments with dramatically increased sensitivity of the
detectors are in progress \cite{gol93}.

The theoretical lifetime $t_l$ of the decaying turbulence
(calculated with $b^{(eff)}=0.55$)
is $t_l= 0.013\tau = 140s$, much larger than the $10 s$ in which the
$\omega(t)$-signal can be measured.
Thus viscous effects, arising
from the $Re$-dependence of $\ceps$ for smaller $Re$, see eq.\
(\ref{eq7}), only become important for a time $t_l \gg 10s$.
So the slight decrease in the measured $\omega (t)$ signal
for $t\approx 10s$ is
not due to them, as one might have thought, but possible due to
the uncoupling of the normal and superfluid components of helium II
(which was used as the fluid in the experiment \cite{smi93a}), as
speculated in \cite{smi93a}.

We summarize our main results. We first calculated the functions
$\ceps (Re_\lambda )$ and $Re_\lambda (Re)$, eqs.\
(\ref{eq3}),
(\ref{eq7}), and (\ref{eq8}), from a variable range mean field theory
\cite{eff87}, which goes far beyond dimensional analysis. We then applied
our results to decaying turbulence, highlighted by the
expression for  the time dependence of $\omega (t)$, eq.\
(\ref{eq16}). All results are in good agreement with experiment.
To even improve the agreement, future work has to be done to
embody non universal boundary effects in this approach.
A way to do so is to introduce $b^{(eff)}$ instead of $b$ in eq.\
(\ref{eq5}).
Alternatively, one could get rid of the boundary effects
by calculating a high passed filtered velocity field from the
experimental data, so that the non universal effects are
filtered out.

\vspace{1.5cm}
\noindent
{\bf Acknowledgements:}
I am grateful to N.\ Goldenfeld, who encouraged me to start this work.
Moreover, I also thank
P.\ Constantin, Ch.\  Doering,
B.\ Eckhardt, A.\ Esser, Th.\ Gebhardt,
L.\ Kadanoff, P.\ Olla, and in particular
S. Grossmann for helpful comments.
Michael Smith kindly supplied me with the
experimental data of Fig.\ 3.
Support by a NATO grant,
attributed by the Deutsche Akademische Austauschdienst (DAAD),
and by DOE is kindly acknowledged.

\newpage

\centerline{\bf Figures}
\begin{figure}[htb]
\caption[]{
The $Re_\lambda$-dependence of $\ceps/\cepsi$ according to eq.\
(\ref{eq8}) with $b=8.4$. Also shown is Sreenivasan's collection of
experimental data
\cite{sre84}. The inset shows $Re_\lambda (Re)$ according
to eqs.\ (\ref{eq3}) and
(\ref{eq7}). The crossover from the laminar scaling $Re_\lambda \propto
Re$ to the turbulent range scaling $Re_\lambda \propto \sqrt{Re}$ takes
place at $Re_{CO}=30$, $Re_{\lambda ,CO} = 22$.
}
\label{fig1}
\end{figure}

\begin{figure}[htb]
\caption[]{
Time dependence of the Taylor-Reynolds number $Re_\lambda (t)$ for
freely decaying turbulence with fixed outer length scale  for different
$Re_0 = 10^7, \quad 10^6, \quad 10^5, \quad 10^4$ ($Re_{\lambda ,0}=
1.58 \cdot 10^4, \quad 4.98\cdot 10^3 \quad
1.58 \cdot 10^3, \quad 498$), left to right, cf.\ eqs.\
(\ref{eq3}),
(\ref{eq12}), and
(\ref{eq13}) with $b=8.4$.
The dashed line denotes the lifetime  $t_l$
i.e., the time, when viscous effects
start to dominate,
see text.
The inset shows
the $Re_{\lambda ,0}$-dependence of the lifetime $t_l$ of the
turbulence, calculated from  eqs.\ (\ref{eq18}),
(\ref{eq12}), and (\ref{eq3}) with $b=8.4$.
}
\label{fig2}
\end{figure}

\begin{figure}[htb]
\caption[]{
Time dependence of the mean vorticity $\omega (t)$
from Smith, Donelly, Goldenfeld, and Vinen's
experiment \cite{smi93a} for
$Re_{\lambda, 0} = 634$, solid line.
In that experiment $L=1cm$, $\nu= 8.97 \cdot 10^{-5} cm^2/s$,
so that $\tau = L^2/\nu = 1.11 \cdot 10^4 s$. The dashed line is our
prediction for $\omega (t)$
according to eq.\
(\ref{eq16})
with $b^{(eff)}=0.55$.
The corresponding lifetime $t_l = 0.013\tau = 140 s$ is given, too.
}
\label{fig3}
\end{figure}

\newpage


\end{document}